\documentclass[12pt]{article}
\usepackage{amssymb,epsf,color}
\usepackage{graphicx}
\def\lsim{\mathrel{\rlap {\raise.5ex\hbox{$ < $}}
{\lower.5ex\hbox{$\sim$}}}}
\def\gsim{\mathrel{\rlap {\raise.5ex\hbox{$ > $}}
{\lower.5ex\hbox{$\sim$}}}} 
\def\sqr#1#2{{\vcenter{\vbox{\hrule height.#2pt

        \hbox{\vrule width.#2pt height#1pt \kern#1pt

           \vrule width.#2pt}

        \hrule height.#2pt}}}}

\def\lsim{{\displaystyle
{{\raise-8pt\hbox{$ <$}}
\atop{\raise5pt\hbox{$\sim$}}}}}
\def\gsim{{\displaystyle
{{\raise-8pt\hbox{$ >$}}
\atop{\raise5pt\hbox{$\sim$}}}}}
\def\slsim{{\displaystyle
{{\raise-8pt\hbox{$\scriptstyle <$}}
\atop{\raise5pt\hbox{$\scriptstyle \sim$}}}}}
\def\sgsim{{\displaystyle
{{\raise-8pt\hbox{$\scriptstyle  >$}}

\atop{\raise5pt\hbox{$\scriptstyle \sim$}}}}}

\newskip\humongous \humongous=0pt plus 1000pt minus 1000pt

\newcommand{\sumpf}[0]{\sum_{(H^{\rm f},G^{\rm f})}\! \! \! \!
{\raise
4pt
\hbox{$'$}}\,}

\newcommand{\sump}[0]{\sum_{(H,G)}\! \! {\raise 4pt \hbox{$'$}}\,}

\def\bs{\begin{subequations}}
\def\es{\end{subequations}}

\catcode`\@=11
\newcount\hour
\newcount\minute
\newtoks\amorpm

\hour=\time\divide\hour by60
\minute=\time{\multiply\hour by60 \global\advance\minute by-\hour}
\edef\standardtime{{\ifnum\hour<12 \global\amorpm={am}%
        \else\global\amorpm={pm}\advance\hour by-12 \fi

        \ifnum\hour=0 \hour=12 \fi
        \number\hour:\ifnum\minute<10 0\fi\number\minute\the\amorpm}}
\edef\militarytime{\number\hour:\ifnum\minute<10 0\fi\number\minute}
\def\draftlabel#1{{\@bsphack\if@filesw {\let\thepage\relax
   \xdef\@gtempa{\write\@auxout{\string
      \newlabel{#1}{{\@currentlabel}{\thepage}}}}}\@gtempa
   \if@nobreak \ifvmode\nobreak\fi\fi\fi\@esphack}
        \gdef\@eqnlabel{#1}}
\def\@eqnlabel{}
\def\@vacuum{}
\def\draftmarginnote#1{\marginpar{\raggedright\scriptsize\tt#1}}
\def\draft{\oddsidemargin -.2truein
        \def\@oddfoot{\sl preliminary draft \hfil
        \rm\thepage\hfil\sl\today\quad\militarytime}
        \let\@evenfoot\@oddfoot \overfullrule 3pt
        \let\label=\draftlabel
        \let\marginnote=\draftmarginnote
   \def\@eqnnum{(\theequation)\rlap{\kern\marginparsep\tt\@eqnlabel}%
\global\let\@eqnlabel\@vacuum}  }
\def\subequations{\refstepcounter{equation}%
  \edef\@savedequation{\the\c@equation}%
  \@stequation=\expandafter{\theequation}
  \edef\@savedtheequation{\the\@stequation}
  \edef\oldtheequation{\theequation}%
  \setcounter{equation}{0}%
  \def\theequation{\oldtheequation\alph{equation}}}

\def\endsubequations{\setcounter{equation}{\@savedequation}%
  \@stequation=\expandafter{\@savedtheequation}%
  \edef\theequation{\the\@stequation}\global\@ignoretrue
  \vspace*{-12pt} \\}

\def\bs{\begin{subequations}}
\def\es{\end{subequations}}
\relax
\def\thefootnote{\fnsymbol{footnote}}
\def\be{\begin{equation}}
\def\ee{\end{equation}}
\def\ba{\begin{eqnarray}}
\def\ea{\end{eqnarray}}

\def\ee{\end{equation}}
\def\bea{\begin{eqnarray}}
\def\eea{\end{eqnarray}}
\def\nn{\nonumber}

\newcommand{\uarrw}[0]{\mathrel{
{\raise.5ex\vbox{\hrule width 1cm}\hskip-6pt\rightarrow}}}
\def\thebibliography#1{%
\vskip 0.5cm \centerline{\bf References}
\list{%
[\arabic{enumi}]}{\settowidth\labelwidth{[#1]}
\leftmargin\labelwidth
\advance\leftmargin\labelsep
\usecounter{enumi}}
\def\newblock{\hskip .11em plus .33em minus .07em}
\sloppy\clubpenalty4000\widowpenalty4000
\sfcode`\.=1000\relax}

\renewcommand{\theequation}{\arabic{section}.\arabic{equation}}

\renewcommand{\section}{\setcounter{equation}{0}\@startsection%
{section}{1}{0mm}{-\baselineskip}{0.5\baselineskip}%
{\normalfont\normalsize\bfseries}}

\renewcommand{\subsection}{\@startsection%
{subsection}{2}{0mm}{-\baselineskip}{0.5\baselineskip}%
{\normalfont\normalsize\slshape}}

\renewcommand{\subsubsection}{\@startsection%
{subsubsection}{2}{0mm}{-\baselineskip}{0.5\baselineskip}%
{\normalfont\normalsize\slshape}}

\begin{document}
%
%
\renewcommand{\theequation}{\arabic{section}.\arabic{equation}}
\begin{titlepage}
\begin{flushright}
\end{flushright}
\begin{centering}
\vspace{1.0in}
\boldmath

{ \large \bf A note on the gravity screening in quantum systems}

\unboldmath
\vspace{1.5 cm}

{\bf Andrea Gregori}$^{\dagger}$ \\
\medskip
\vspace{3.2cm}
{\bf Abstract} \\
\end{centering} 
\vspace{.2in}
We discuss how, in  the theoretical scenario presented in \cite{assiom-2011}, 
the gravity screening and the gravity impulse
which seem to be produced  under certain conditions by high temperature superconductors
are expected to be an entropic response to the flow of
part of the system into a deeper quantum regime.

\vspace{6cm}

\hrule width 6.7cm
\noindent
$^{\dagger}$e-mail: agregori@libero.it

\end{titlepage}
\newpage
\setcounter{footnote}{0}
\renewcommand{\thefootnote}{\arabic{footnote}}

\section{Introduction}
\label{intro}

Although not a new branch of research, the physics of superconductors
appears the more and more to provide us with a source of
experimental and theoretical challenges,
calling for a better understanding of quantum phenomena.
In particular, this is true for high temperature superconductors.
In a recent work, we proposed the existence of a relation between
the critical temperature of superconduction
and the energy distribution of the geometric structure, i.e. the geometry 
of the superconductor intended in the
sense of General Relativity~\cite{spc-gregori}. 
The theoretical ground of this idea
is that, at a more fundamental level than the one of quantum electrodynamics,
quantization of the geometry of space cannot be neglected. Its effects, that
can be generally referred to as "quantum gravity", do not simply
involve quantization of the propagating gravitational field (i.e. 
roughly speaking treating the graviton
in a similar way to the photon), but imply quantization of the geometry
of space itself. As a result, one obtains an effective dependence of the
quantum delocalization, or, if one prefers, of the Planck constant as it enters in the Uncertainty Principle, on the geometry of space. 
Taking this into account, we were able to compute several critical temperatures of
high temperature superconductors, as a pure theoretical prediction
from the analysis of their lattice structure. 
In this note, we want to consider some other important aspects of the physics
of high temperature superconductors, which are
the object of (controversial) experimental investigation, namely the detection of some
kind of gravity screening produced by rotating superconductors, or gravity-like
impulse produced by a superconductor undergoing an electric discharge
\cite{Podkletnov1997}, \cite{Podkletnov:2001gr}, \cite{PodkletnovModanese2002}. Attempts to explain these effects in the light of
a quantum gravity theoretical framework have been proposed 
in~\cite{Modanese:1995tx},
\cite{Modanese:1996zm}, \cite{ModaneseJunker2009}.
Here we want to discuss how all these effects are a natural prediction
of the theoretical scenario described in~\cite{assiom},
updated in~\cite{assiom-2011},
namely the same theoretical framework in which critical temperatures
have been investigated in \cite{spc-gregori}.
This discussion will allow us to point out also some important aspects
of the quantum gravity scenario implied in this framework, in particular
about how all forces, and the equivalence principle at the ground of General Relativity
itself, arise from an entropic principle.

\section{A short summary of the theoretical set up}
\label{setup}

In Refs.~\cite{assiom-2011} we have introduced a physical scenario 
describing a universe ruled by a dynamics which embeds quantum mechanics
and general relativity, to which it reduces under appropriate conditions.
The basic idea is that the universe is not a particular
configuration of whatever kind, among those predicted within the
phase space of a certain theory, but the whole of all possible codes of
information, time-ordered according to the inclusion of sets,
that we interpret as configurations of energy distributed along the space.
In some sense, this can be viewed as the generalization of the idea of the
Feynman path integral, according to which the time evolution of
a quantum system is given by the weighted sum over all 
trajectories~\footnote{For a discussion of the relation to
the Feynman path integral, see~\cite{assiom-2011}, section~6.4.}.
Let us here briefly summarize the main lines of the set up.
Consider the set $\Phi = \{  \psi (N) \}$, $N \in {\cal N}$,
of all the distributions of an amount $N$ of 
energy units along a target vector space, of any possible dimension, for any $N$.
The subsets $\Phi (N) \equiv \{  \psi (N) \}$ have a natural ordering
through $N$, because $\forall \psi(N) \in \Phi (N)$ 
$ \exists \psi (M) \in \Phi (M)$, $M > N$, such that $\psi (N) \subset \psi (M)$.
$N$ can therefore be identified
with the "time". 
The partition function of the universe at time ${\cal T} = N$ is:
\be
{\cal Z}_{N} = \sum_{\psi(N)} {\rm e}^{S(\psi)} \, ,
\label{zsum}
\ee
where $W(\psi) \equiv \exp S(\psi)$ is the weight of a configuration $\psi$
in the phase space, $S$ being as usual the entropy. 
$N$ not only works as "time", but is also the total energy of the universe at time $N$, implying the identification $E = {\cal T}$.
The distributions of energy units
in a discrete vector space of cells can be viewed as assignments of 
"geometry" given by occupation codes 
(binary codes, of the type full/empty), and therefore seen as 
corresponding to codes of information. 
Through this correspondence,
one can see that different geometries correspond 
to different discrete groups of symmetry.
The classification of these spaces amounts therefore to
a classification of all possible discrete (and in general not simple) groups. 
This correspondence allows to make
the key observation that different codes
of information, i.e. different distributions of energy,
correspond to different groups of symmetry,
and therefore have a different weight in
the phase space of all the configurations, 
the latter being related to the volume of the corresponding group. 
Expression~\ref{zsum}
says that the universe looks the most like the
geometry which is realized in the highest number
of equivalent ways, i.e. the one which has the highest weight, 
or the highest entropy.

At any energy, and time, $E \sim {\cal T}$,
the dominant configuration implied by \ref{zsum} is a three-sphere
of radius $R \sim {\cal T}$,
i.e. a universe predominantly consisting of a space with three dimensions,
and the curvature of a three-sphere $\sim 1 / {\cal T}^2$.
Its weight $W$ in the phase space $\Phi$ is proportional to $\exp N^2$.
The entropy is therefore $S \sim N^2 \sim {\cal T}^2 \sim R^2$.
The dominant configuration can be viewed as the one describing
the "classical" part of the geometry of the universe.
The contribution to any mean value as due to the configurations
different from the dominant one amounts to a
"smearing" of the classical value of the order of the Heisenberg Uncertainty.
Quantum mechanics arises in this framework as a way of
implementing this undefinedness, and the uncertainty principle expresses the fact
that any observable in the three-dimensional world is indeed
just the average configuration of something which receives contribution
from any configuration, in any dimension. The physical world is only
in the average three-dimensional, and beyond a certain degree of 
accuracy the physical quantities and any degree of freedom 
as is characterized in three dimensions cannot not only be measured, but not
even be defined. The probabilistic interpretation
introduced in quantum mechanics is a way to deal with
the undefinability of any measurable quantity beyond a certain accuracy, 
by embedding it in a consistent mathematical-theoretical framework.
This enables making computations, and 
predictions up to a certain degree of accuracy, while keeping
under control the "unknown".

\section{Forces from entropy}
\label{forcentropic}

In this framework, the universe is a staple of configurations, 
and what we call dynamics, and usually describe in terms of forces and interactions,
is a parametrization of the changes which occur in the geometry
produced by this staple of configurations. Our distinction 
of interactions and forces into classical and quantum mechanical ones 
regards the level at which we want to consider the physical world, i.e. the degree of approximation we introduce when giving our description of
physical phenomena, and which depends on what kind
of configurations we decide to neglect. 
Roughly speaking, classical physics basically corresponds to considering just
the most entropic configurations. Since configurations remote in the phase space
have weights which are exponentially suppressed as compared to the most entropic
ones, with an abuse of language we often indicate classical values as "mean" values. 
The most entropic of all the configurations
describes a sphere in three dimensions., i.e. a completely homogeneous distribution 
of energy. Were this the only configuration of the universe, the only 
force existing would be a gravitational force acting in a way
to distribute homogeneously energy along this space.
However, the contribution of less entropic, and correspondingly
less homogeneous, configurations is responsible for
the formation of "clusters". It is thanks to this that we can perceive
gravity as a force which is locally attractive, i.e. which acts in a way to break 
the complete homogeneousness of the energy distribution.
Including the more and more configurations in our approximation
of the universe, up to virtually all the configurations, leads to
a quantum mechanical description of physics. This includes
also other types of interactions besides gravity, introduced in order
to parametrize a more intricate structure in which certain types of energy clusters,
that we call particles, do not simply attract each other,
but interact in a more complicated way.

In this context, what we call gravitational attraction is a way
of theoretically parametrize in terms of forces the fact that two
objects tend to go closer to each other, because in this way the
entropy of the space around each one, 
corresponding to the entropy of its energy distribution, 
gets increased by an amount corresponding to the entropy of the energy
distribution of the other object.
Of course, in the sum over all configurations \ref{zsum},
also other motions are included. Indeed, \emph{all} possible
motions are counted. However, they give an exponentially
suppressed contribution to whatever mean value of observable
quantity, so that it makes sense, at least approximately, 
to speak of "motion tending to increase entropy" in classical terms.  
Indeed, in this theoretical framework \emph{all} forces, not just the gravitational
one, are entropic.

\section{Effective $\hbar$ and quantum delocalization}

In the traditional approach to quantum mechanics,
the Planck constant sets by convention the size of the quantum uncertainty, namely, the "normalization" of the Heisenberg uncertainty relations. On the other hand, in this
way it naturally sets also 
a unit of conversion from momentum to space, or from energy to time.
In our framework, the unit of energy/time conversion 
is a quantity \emph{independent} on the scale of quantum delocalization. 
In our case, the "canonical" expression of the uncertainty relation:
\be
\Delta E \Delta t \geq {\hbar \over 2} \, ,
\label{DeDtC}
\ee 
encodes the fuzziness in mean values as given by the contribution
of all the configurations \emph{out of the most entropic one}, which corresponds 
to a universe with the ground, classical curvature $\sim 1 / {\cal T}^2$.
In this framework, empty space
does not exist: the minimal curvature is the average curvature
of the three-sphere geometry of the universe at age ${\cal T}$,
$1 / {\cal T}^2$ is what corresponds to
the curvature of the "empty space". One can view this as
the condition which corresponds to considering only the cosmological term of
the Einstein's equations, neglecting all other contributions, that come
from the local distribution of masses and energies of particles and fields.
\ref{DeDtC} is therefore also the uncertainty relation ideally ruling the behavior
of what in usual terms is the "free" electron, or the free electromagnetic field,
indeed conditions in some cases well approximated, but
rigorously never realized in practice.
When the physical system  
presents a more complicated geometry, this approximation is
no more valid, and the relation~\ref{DeDtC}
must be corrected. The reason is that the configuration
describing the system is in this case a superposition of configurations 
with an average entropy lower than the highest one, the one of the
"vacuum" of the universe, the three-sphere; the relative weight
of less entropic configurations compared to the one of the system is 
therefore higher (see sections~2.7 and 3 of \cite{assiom-2011}).  
As a consequence, geometrically remote systems are more quantum-delocalized.
A typical case of physical system in which the higher amount of quantum delocalization
shows out in a clearly detectable way is the one of superconductors.

As discussed in Ref.~\cite{assiom-2011}, 
in the approximation of neglecting configurations extremely delocalized, and therefore
also very remote in the phase space, one can 
factor out from the expression of the
weights in the phase space overall space volume factors, 
which can be considered to be the same
for all the configurations, and consider the contribution to the
weight due just to the internal symmetry of a configuration. Under this approximation,
the entropy of the configuration corresponding
to the three-sphere scales as $S_0 \sim {\cal T}^2$.
Factoring out common volume factors, and approximating sums with integrals,
one can therefore
write an effective partition function of the universe in the form:
\be
{\cal Z}_{\rm eff} \sim \int_{S = S_0}^0 d S \, {\rm e}^{S} \, .
\label{Zeff}
\ee 
Isolating the dominant term in the integrand, $\exp S_0$, one obtains a correction
of the order $\sim (1 / S_0) \exp S_0$. Remembering that $S_0 \sim {\cal T}^2$,
and identifying ${\cal T}$ with the duration $\Delta t$
of the experiment of measurement, in this case corresponding to
the existence of the universe itself, one obtains that,
during the interval $\Delta t$, the correction to the energy of the universe
$E \sim \Delta t$, is of the order of:
\be 
\Delta E ~ \sim ~ E \, \Delta {\cal Z}_{\rm eff} \; \approx \; {\Delta t \over (\Delta t)^2}
~ = ~ {1 \over \Delta t} \, .  
\label{DeDtG}
\ee
This equality
is the lower bound of the ordinary form of the Heisenberg Uncertainty.
Indeed, the quantum uncertainty comes from an effective linearization
operated on the phase space when isolating a configuration with
highest entropy: the uncertainty is treated as a perturbation of this
configuration. This leads to a dependence of the amount of
delocalization on the entropy, rather than on the
weight itself of the configuration in the phase space. 

Let us now consider cases different from the measurement of the energy of the
universe itself: let us consider local experiments. 
A local experiment corresponds to a subset of the universe. In this case, the
classical shape around which to expand in order to find the quantum corrections
is a superposition of geometries, and one should better speak in terms of "mean"
entropy and average geometry $\overline{S}$
(see~\cite{spc-gregori} for a detailed discussion). In any case, in general
$\overline{S} < S_0 \sim (\Delta t)^2$, and
expression~\ref{DeDtG} becomes:
\be
\Delta \overline{E} ~ \sim ~ {\Delta t \over \overline{S}} \; > \, {1 \over \Delta t} \, .
\ee
Indeed, one obtains:
\be
\Delta \overline{E} ~ \sim ~ \left( {\overline{S}  \over S_0} \right)^{-1} \times {1 \over \Delta t} \, .
\ee
The delocalizations of two systems with different entropy stay therefore
in ratio:
\be
{\Delta t_i \over \Delta t_j} \, = \, {\Delta X_i \over \Delta X_j} ~ = ~
{\overline{S}_j \over \overline{S}_i} \, .
\label{DiDj}
\ee
On the other hand, since what we are considering are subsets of the whole
universe, involving an amount of energy, and entropy, much smaller
than the one of the universe, $S({\rm experiment})
\ll S({\rm environment})$, we can decompose the weight of
the whole configuration of the universe which includes our experiment as:
\ba
W & = & W({\rm environment}) \times W({\rm experiment}) ~
= ~ {\rm e}^{S({\rm env.}) \, + \, S({\rm exp.})} \nn \\
& \approx & W({\rm env.}) \times 
\left(1 + {S({\rm exp.}) \over S({\rm env.}) } \right)
\, + \, {\cal O}\left[ \left( {S({\rm exp.}) \over S({\rm env.}) } \right)^2 \right] \, .
\label{Wee}
\ea
Isolating the system corresponding to the experiment
corresponds to factoring out the contribution of the environment,
i.e. discarding the first term in the r.h.s. of \ref{Wee}. This
allows to approximate the ratio of the delocalizations of different systems as:
\be
{\Delta_i \over \Delta_j  } ~ = ~ {S_j \over S_i} ~ \approx ~ {W_j ({\rm exp.}) \over 
W_i ({\rm exp.})} \, .
\label{DDWW}
\ee
On the other hand, as described in~\cite{spc-gregori}
the ratio of the weights of two configurations can be written as
the ratio of the volumes of the symmetry group of the
energy distributions representing the two configurations.
Thanks to the factorization \ref{Wee}, this property transfers also
to the subgroups of the symmetry of the whole configuration which
correspond to the experiment, $G_i$ and $G_j$. The ratio \ref{DDWW}
can therefore be written as:
\be
{\Delta_i \over \Delta_j  } ~ = ~ { ||G_j|| \over ||G_i||} \, .
\label{DDGG}
\ee
This holds both for the weight of single configurations, and for the mean weight
in the case one deals with a superposition of configurations, as 
is the case of concrete local experiments.
If the concept of mean weight in the case of a superposition of configurations
doesn't sound something unnatural, it is less obvious that one can speak of 
symmetry groups for a superposition. However, as discussed 
in~\cite{spc-gregori}, any mean weight can be approximated with
the weight of a discrete group, which is then uniquely identified. This can be
assumed to be the weight entering in expressions like~\ref{DDGG}. 

Expression~\ref{DDGG} says that
the higher is the degree of inhomogeneity of the physical configuration,
the more its description enters into a deep quantum regime. 
In this theoretical framework, the Uncertainty
Principle parametrizes in the form of a bound on the fuzziness of observables
the fact of being the world a staple of configurations. 
However, the degree of fuzziness (or quantum delocalization) depends on
how remote is the configuration one wants to consider.
According to this interpretation, traditional quantum mechanics
considers just the "first level" of quantum delocalization, the one
in which the details of the geometry of the energy distribution of a microscopic
system are neglected. It is therefore not adequate for the description of
systems in which this approximation is not valid. Taking into account
the effects due to the geometry of space leads us to a regime
of quantum geometry. We can call this a regime of quantum gravity, provided
we intend that this does not simply mean 
the description of gravitational interactions in terms of the propagation of gravitons,
quantized in a way possibly similar to, or reminiscent of, the quantization
of the electromagnetic field in terms of photons. Quantization
of the geometry means, and implies, much more. In particular,
it implies that the degree of quantum delocalization of wavefunctions
depends on the geometry. One can view this as the result
of the fact that quantization of geometry roughly means the introduction of
a metric $g_{\mu \nu}$ which depends on $\hbar$; 
considering instead as a renormalization prescription the geometry of a system 
to be fixed, and thereby inverting the relation $g_{\mu \nu} =
g_{\mu \nu}(\hbar)$, this equivalently means the introduction
of an effective metric-dependent Planck 
constant: $\hbar = \hbar(g_{\mu \nu})$.

\section{High temperature superconductors}
\label{HTSC}

As discussed in~\cite{spc-gregori},  
considering this effect allows to justify 
higher critical temperatures for superconductors with a complex
lattice structure, and predict their dependence on
the geometry of the lattice for a wide class of superconductors,
on the base of the same mechanism that allows to explain
superconductivity at low temperature: the formation
of Cooper's pairs as in the BCS theory. Simply, the critical condition
for superconductivity regime,
i.e. having an appropriate number of electrons with an appropriate
amount of delocalization, is attained at a higher energy because
electrons need a lower localization in the space of energies and momenta
to attain the same degree of delocalization in space.
The ratio of the critical temperatures of two superconductors
is therefore expected to be related
to the ratio of their quantum delocalizations:
\be
{T^i_c \over T^j_c} ~ = ~ { \Delta_i \over \Delta_j  } \, ,
\ee 
in turn depending, from~\ref{DiDj},
\ref{Wee} and \ref{DDWW}, on the
entropy, and approximately the weight, of the respective geometries:
\be
{T^i_c \over T^j_c} ~ \approx ~ {W_j \over W_i} \, .
\label{TiTj}
\ee
Comparing this with expression~\ref{DDGG} 
we see that~\ref{TiTj} can be written as:
\be
{T^i_c \over T^j_c} ~ = ~ {|| G_j || \over || G_i ||} \, .
\label{TTGG}
\ee
In Ref.~\cite{spc-gregori} the ratio of these volumes
was approximated by the ratio of the average space gradients of energy:
\be
{T^i_c \over T^j_c} ~ \approx ~ {\int_{a_i} |\nabla E_i | \over 
\int_{a_j} |\nabla E_j|} \, ,
\label{TTEE}
\ee
where $a_i$ and $a_j$ are the characteristic lengths 
(in general corresponding to the lattice length) of the two
superconductors. 
Roughly speaking, this can be understood by considering
that the degree of symmetry measures how smooth the
energy distribution is.
From expression~\ref{TTEE} it has been possible to
predict the critical temperature of whole families of high temperature
superconductors from the analysis of their crystalline structure,
finding a remarkable agreement with the temperatures experimentally measured.

\section{Black holes}

\begin{figure}
\centerline{
\includegraphics[width=12cm]{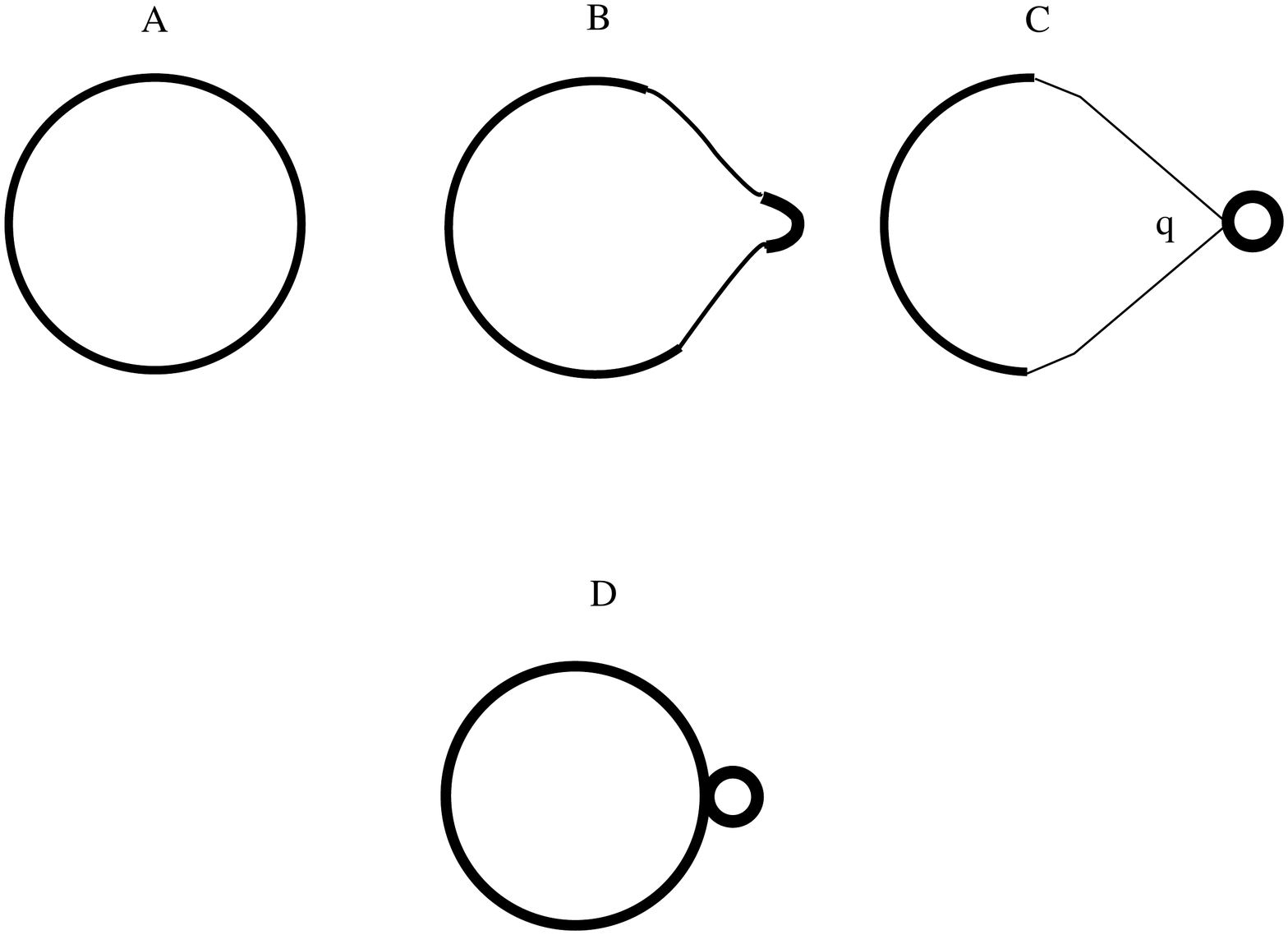}
}
\vspace{1.5cm}
\caption{The process of creating a black hole within the universe: from the
three-sphere universe (A) we move energy units, thereby locally flattening the space,
to create a bump (B), and then
a throttling, squeezed to a point in q, to factor out the smaller three-sphere (C).
The different degrees of energy density are indicated through a different thickness of
the line. The symmetry of the configuration (C) is always lower than that
of the configuration (D), the product of two three-spheres,
built with the same amount of energy.}
\label{throttle}
\end{figure}

A black hole 
has an entropy equivalent to the one of a three-sphere
with radius proportional to the Schwarzschild radius. 
Considered as a standalone object, it is therefore a highest symmetric, highest
entropic space. However, when considered as inserted in a larger universe,
one can see that black holes are the most singular configurations
of the three-dimensional space.
The most entropic configuration of the universe is the one
in which the energy units are uniformly distributed to form the geometry of a 
three-sphere. Configurations which contain clusters of energy
are less symmetric, and therefore less entropic: we can in fact think to form
energy clusters out of the most homogeneous energy distribution, by moving, 
one after the other, a certain amount of energy units from their initial position toward 
certain regions of agglomeration. While doing this, we clearly reduce the symmetry of the
configuration. In order to form a subspace of the
whole universe which behaves already in itself like a small universe, we must
move energy units to create a bottleneck (a throttling)
till we "throttle" the space in a certain point
(see figure~\ref{throttle}). 
In this way, we obtain a configuration in which
the space is factorized into a black hole times the rest of the universe.
This configuration has an entropy which is at most the one of a small three-sphere
(the black hole) times a three-sphere formed with the remaining energy of the universe.
If the total energy is $N$, and we take out $n$ energy units to form
the black hole, the weight $\exp S^{\prime}$ of the new configuration
satisfies the following condition:
\be
{\rm e}^{S^{\prime}} ~ \lesssim ~ {\rm e}^{n^2} \times {\rm e}^{(N-n)^2} ~ 
\lsim ~ {\rm e}^{-N} \times {\rm e}^{N^2}\, .
\ee
It is therefore exponentially suppressed (at least a factor $\exp -N$)
as compared to the configuration
of highest entropy of the universe at energy, and time, $N$.
As discussed in~\cite{assiom-2011}, configurations
with this weight belong to the bunch of those so remote to
describe a deeply quantum regime. Indeed, from~\ref{Zeff}
one can see that the highest contribution
to the mean value of the energy density in the region of space of the black hole 
comes from configurations with higher overall entropy, which 
however do not describe a 
black hole. The exponential suppression of the weights implies that the average
energy density in the region of the black hole is lower than the
critical one. This in practice means that the black hole does not exist.
According to this theoretical scenario, the only black hole which can exist
is the universe itself. 
For the sake of our present discussion what is important
is however to remark that,
intended as a configuration describing 
a \emph{subset} of the whole universe, a black hole is the less entropic one
among all the possible in three dimensions, and describes a 
limit configuration of deep quantum regime.

\section{Gedankenexperiment}

For our discussion it is not so important
whether black holes can really exist, apart from the universe itself, or not.
We are going to treat them as exemplar cases, useful to the purpose of
understanding real physical cases. What we want to see here is 
what happens when in the universe a region of space undergoes a transition
to a deep quantum regime, that we ideally assume to be a black hole.
As discussed in \cite{assiom-2011}, our theoretical scenario implies,
in its classical limit, general relativity. From a macroscopic point of view, 
in order to investigate some large scale properties of gravity and geometry of space,
not too close to the Schwarzschild horizon,
we can therefore use much of what is known about the metric around a
classical black hole.

Let us consider observing from a point (O) 
an object (A) gravitationally attracted by another one, (B),
that we suppose of mass much larger, so that we can neglect
its motion, consider it at rest and work with masses instead than with reduced masses.
What we see is that object (A) moves with an accelerated motion toward (B).
Let us now suppose to shrink the radius of (B) beyond the Schwarzschild value,
i.e. to the point of making of it a black hole (for the sake of our
discussion, it is not important here to inquire
whether this transformation is physically possible, and, in the case, in which
way and as a consequence of what this can occur).
Owing to the created Schwarzschild singularity in the new metric,
the motion of (A) will now appear to an external observer to
undergo a deceleration, because the time it takes for an object in order to
reach the surface of the black hole diverges. Saying that 
(A) needs an infinite time to get to a point placed at a (seemingly) well defined, and finite,
distance means saying that (A) gets stopped, until it is at rest, at
the Schwarzschild surface. As seen from (O), the situation is therefore that
some kind of force has been created, which opposes to gravity. This new force
has all the properties of gravity, in the sense that it seems to act with twice the strength
on objects with twice the mass of (A). 
It has therefore all the characteristics of a "gravity screening".

The surface at the horizon of a black hole is a limit example
of extremely non-classical, "quantum" geometry. 
However, although concerning a rather unphysical situation,
this thought experiment helps us in understanding
how the presence of a very low-entropic configuration of space
acts on the overall geometry in a way to "create" a kind of gravity screening
force, or, in dynamical cases, anti-gravity impulse.
In our scenario, the dynamics is ruled by the "law" of entropy, in the sense that 
at any time the universe is predominantly given by the configuration of highest
entropy in the phase space. Instead of speaking of "forces", "anti-forces"
and similar concepts, the best way to analyze the behavior of a system 
in response to a certain event, is to compare entropies of configurations.
In order to see whether after the creation of the black hole (A) is further
attracted by (B), or starts to be repelled, we have to see whether
a motion toward (B) leads to an increase of the entropy of the configuration, or to
a decrease. In section~\ref{forcentropic} we have justified the gravitational attraction with
the fact that, by moving toward each other, two objects
reciprocally increase the entropy of their configuration, because in their phase
space, besides the entropy due to the energy distribution of each single object, one
adds also the entropy due to the energy of the other object: the 
"space density of entropy" increases. In falling
toward a black hole, an object enters in a region of space which corresponds
to configurations the more and more remote in the phase space.
The object itself becomes the less and less classical. Its entropy
decreases. When going away
from the black hole, the entropy of the configuration it corresponds to
instead increases. This is in our framework the origin of
the "repulsion" the object appears to experience. 
Indeed, the true effective motion depends
also on all the rest of the universe, to begin with on the closer
environment. In traditional words, this translates in terms
of "initial conditions", "continuity of the motion", acceleration and deceleration, gravitational
attraction of other bodies, etc...
In practice, the balance of entropies doesn't lead in this case to
a net repulsion, but to a decelerated fall of (A) toward (B), to the point that,
before reaching the horizon, i.e. even before being completely stopped in its motion,
(A) looses any property of a classical object, and cannot be anymore
described in the same terms as it was before. The reason of
a deceleration instead of a net escape from the black hole is that,
before becoming a non-classical object, the phase space of (A) still feels
an increase of entropy by falling toward the black hole,
due to the classical part of the energy distribution
of (B) (the "gravitational field" of (B), in classical words).

\section{Gravity screening and gravity impulse in high temperature superconductors}

In our scenario, there is no net separation between classical and quantum mechanical
regime: these two levels of the physical description are sued together in a 
seamless way.
This in particular means that the gravitational screening occurring in
the case of the black hole is not a property specific of black holes, but, 
in a certain amount, is expected to occur in every physical system.
In general, the gravitational attraction does not
depend only on the mass of the object, but is affected also by a certain amount
of screening, depending on the configuration of the object: more
singular (=less entropic) configurations should present a higher degree of screening.
What distinguishes the situation of a so-called classical object 
from a quantum one is that
the entropy of the classical object is very large, to the point that,
in comparison, entropy variations depending on the shape are negligible.
As a consequence, negligible is also the size of the gravity screening as compared to
the gravitational force. 
Quantum systems correspond by definition to less entropic configurations,
and the gravity screening is therefore comparatively higher. They are
therefore good candidates for the detection of this effect.

\begin{figure}
\centerline{
\includegraphics[width=11cm]{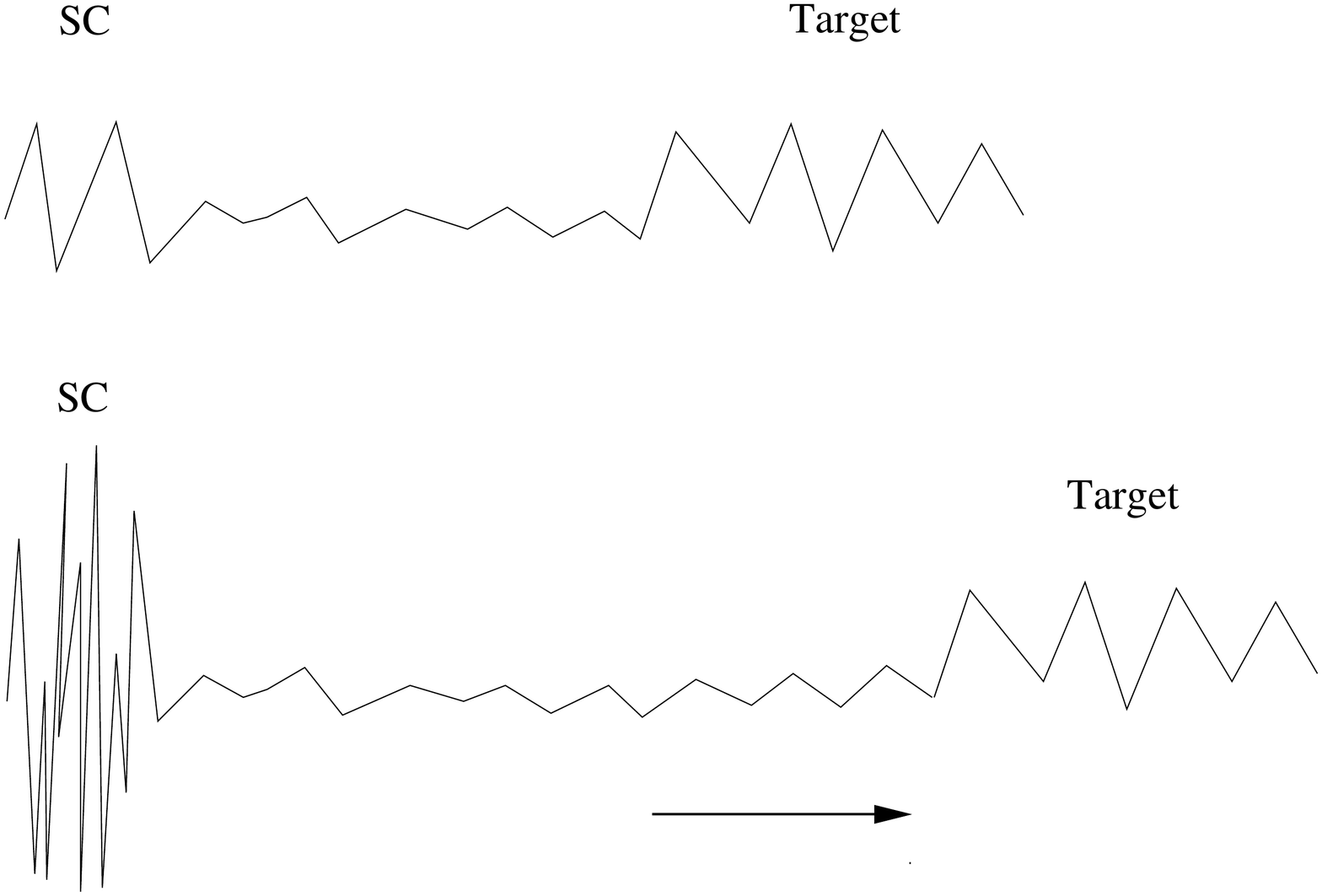}
}
\vspace{1.5cm}
\caption{Increasing the singularity of
the geometry of the configuration (= lowering the entropy) 
in the superconductor leads, as a response of the system, 
to a gravity screening, or gravitational impulse,
tending to push the target massive object further away from the superconductor, in order
to counterbalance the decrease of entropy by producing
a configuration which, on the right of the superconductor, has a lower average
density of energy gradient (= average lower degree of singularity).}
\label{gravimpulse}
\end{figure}

If the case of a black hole is just an ideal case, a very concrete physical case
of region of space which is very remote in the phase space (i.e. much
less entropic than its environment), although not so extremely remote as 
the surface of a black hole, is a superconductor with very complicated
lattice structure, like those considered in section~\ref{HTSC}.
Experiments have been carried out by Podkletnov with a
rotating YBCO disc: under appropriate conditions, a loss of weight is observed
in objects hanging above the superconductor.
In particular, the gravity-screening effect results to be enhanced during transitions
of the configuration of the superconductor. This agrees with our expectation
that what matters for the production of the effect is the creation of a region
with a very singular geometry, i.e. of very low entropy in the phase space.
Such a condition can be attained by considering a superconductor 
with a high gradient of the energy distribution, like the high temperature superconductors, 
in a deeply quantum regime, i.e. below the critical temperature
and with accelerated currents. 
Why does the acceleration correspond to a configuration
of even lower entropy? In section~\ref{forcentropic} we said that in our framework
the dynamics underlying the gravitational attraction is ruled by
the "law" of highest entropy. Since the dominant configuration, the
one which the most contributes to the mean value of the observables,
is the one of highest entropy, at any time one produces somewhere a
change leading to a decrease of the entropy, from an effective
point of view what happens is that the system responds tending
to counterbalance this action by changing its configuration
toward an increase of entropy wherever possible. For instance,
if we create a black hole,
an object falling toward it decelerates because it tends to lower the decrease of entropy
produced by its falling into a region of very low entropy, and so on.
The equivalence principle, i.e. the equivalence of gravitational attraction
and acceleration, in our framework is nothing else than the statement that
producing an acceleration means forcing a system to change its entropy.
In practice, this means that by inducing on it an accelerated motion, we
force the system into a regime of lower entropy, to which it responds with
a tendency to counterbalance this effect by
increasing its entropy as in the case of the gravitational attraction. 
In the case the superconductor,
forcing it to a configuration in which the currents and its whole lattice
structure (its whole geometry) is accelerated means that we produce on it
a configuration of even lower entropy, therefore even more remote
in the phase space. From~\ref{TiTj}--\ref{TTEE} we
see that at all the effects it is like having increased
the average energy gradient. 
Creating a region of very low entropy produces 
therefore a "gravitational response"
similar to the one of the black hole. Indeed,
such a kind of response is to be expected
for any kind of similar situation, and is not in principle
related to the fact of having a superconductor: important is
the creation of a region with a deeply quantum behavior.
The response can occur everywhere in the environment: 
the only requirement is that it most efficiently compensates
the decrease of entropy. The direction in which it is detected depends
therefore on conditions such as the symmetries of the problem and of the
environment, which state where a change of configuration is possible, and
how is the best (= most entropic) way.
In the case of a superconductor rotating around the $z$-axes,
since on the plane of the rotation the superconductor is constrained
(that means forces are balanced and forbid any deformation)
the only change of the system
can occur along the direction of the rotation axes, i.e. the vertical axes, 
orthogonal to the superconducting layers. 
The result is a
screening of the gravitational attraction acting on objects hanging over
the superconductor.
As long as we neglect border effects, on the plane of the disc
the layers can approximately be considered of infinite extension and the problem
reduces to a one-dimensional one.
The region of space expected to be affected
by the screening is therefore a cylinder, a column
with the same diameter of the superconductor. The range of the effect,
namely how far should it be detectable,
depends once again on the geometry of the physical system and its environment, i.e.
on how heavy can be the changes in the geometry close to
superconductor, counterbalancing the reduction of entropy on the superconductor.
We expect that in the ideal case of absolutely "rigid" system, the screening can be effective even very far from the superconductor. 
Indeed, like in the case of the black hole, this effect is 
of gravitational type.
As one can also see in figure~\ref{gravimpulse}, the strength
of the screening depends on the amount of reduction of entropy the repulsion
is aimed to compensate. When this quantity is fixed, the amount of the
response \emph{at the level of change in the geometry} is fixed.
This means: what is fixed is the acceleration, not its translation in terms
of force. Let us see this more in detail.
Accelerating the superconducting disc means
making its configuration to weight less in the phase space,
by an amount approximately given by the ratio
of the symmetry groups after and before the acceleration:
\be
{W^{\prime}_{\rm SC} \over W_{\rm SC}} ~ \approx ~ 
{|| G^{\prime}_{\rm SC}|| \over || G_{\rm SC} ||} 
\, .
\label{wpwSC}
\ee
The system superconductor+environment+probe will re-act in a way
to compensate this loss of weight by a move intended to
locally increase the weight somewhere else.
The only action the system is free to do is to change the weight of the probe, i.e.
push it away from the superconductor
(see figure~\ref{gravimpulse}), 
in order to increase its entropy by producing a smoother 
(more symmetric) configuration,
as the result of a decrease of the mean value of the energy gradient. 
Owing to the factorization of the phase space,
this compensation will be obtained by changes leading to a ratio of local weights
inverse to~\ref{wpwSC}:
\be
{ W^{\prime}_{\rm probe} \over W_{\rm probe}}  ~ = ~ 
{ W_{\rm SC} \over W^{\prime}_{\rm SC}} 
~ \approx ~ 
{|| G^{\prime}_{\rm env.+probe}|| \over || G_{\rm env.+probe} ||} 
~ \approx ~ {\int | \nabla E_{\rm env. + probe}| \over 
\int | \nabla E^{\prime}_{\rm env.+probe} |}\, .
\label{Wprobe}
\ee
Being the probe a rigid body, this implies a fixed
change of the \emph{relative} gravitational weight of the probe, the same 
for any probe, as fixed by the ratio~\ref{wpwSC}.
Saying that we have a fixed amount of relative change
precisely means that the amount of change is fixed once it is
divided by the mass of the probe. That means, what is fixed is not the
force but the field strength, precisely what happens in the case of the
screening produced by the sudden creation of a black hole:
a gravitational screening.
Of course, this is a very simplified representation of the physical situation, which
is composed by many other pieces. For instance, the probe hangs from 
a balance, so we should include in our considerations also the balance, etc...
However, a little thought about the fact that all forces are at equilibrium
should convince the reader that this simplification catches the 
main point about what is happening. 
Giving a quantitative prediction of the screening effect is quite hard, but we can at least attempt to estimate a \emph{relative} effect. What we expect is that it should
be relatively simple to predict the change in the strength of the gravity screening
effect when substituting a superconductor with another one. The difficulty in the
theoretical analysis is in the fact that, differently from the situations
considered in \cite{spc-gregori}, here we are going to consider
superconductors in an accelerated state. Namely, we are going to modify
the entropy of their configuration, which now does no more depend roughly
on just their lattice structure. 
Nevertheless, with a certain degree of approximation we can
assume that, precisely owing to the implementation of the equivalence 
principle, the acceleration imposed on the superconductor is
theoretically equivalent to a modification of the close environment of the
superconductor, as produced by the presence of an object of large mass.
This equivalence should be regarded only as a way of
parametrizing the changes in the entropy of the configuration
of the superconductor plus its close environment, i.e. not in order
to treat the whole experiment as being subjected to a modification
equivalent to introducing the presence of an object of large mass, something
that would gravitationally attract also the probe.
We expect it to be reasonable to write 
the contribution of the rotation in terms of an equivalent energy gradient term
$\Delta_{\rm acc.} $, to be added to the energy gradient of the superconductor:
\be
\int_a | \nabla E | ~ \to ~ \int_a | \nabla E | \, + \, \Delta_{\rm acc.} \, ,
\ee
\be
\Delta_{\rm acc.} ~ \equiv ~  "\left( \int | \nabla E |_{\equiv \, 
{\rm acc.}} \right)" \, .
\ee
The weight in the phase space
of the superconductor subjected to a rotational acceleration should
then be: 
\be
W^{\prime}_{\rm SC} ~ \approx ~ \propto  {1 \over \int_a | \nabla E | + 
\Delta_{\rm acc.}} \, .
\ee
The ratio of the weights of two different materials
subjected to the same acceleration should then be:
\be
{W^{\prime}_{\rm SC} \over W^{\prime}_{\rm SC^{\prime}} }
~ \approx ~
{  \int_{a^{\prime}} | \nabla E |^{\prime} + 
\Delta_{\rm acc.}  \over   \int_a | \nabla E | + 
\Delta_{\rm acc.}} 
~ = ~
{W_{\rm SC^{\prime}} \left( 1 \, + \, {\Delta_{\rm acc.} \over
W_{\rm SC^{\prime}}} \right)
\over
W_{\rm SC} \left( 1 \, + \, {\Delta_{\rm acc.} \over
W_{\rm SC}} \right)
} \, .
\label{WDE}
\ee
From \ref{Wprobe} we obtain that the
weight of the probe should be:
\be
{W_{\rm probe} \over W^{\prime}_{\rm probe} }
~ \approx ~ \, {W^{\prime}_{\rm SC^{\prime}} \over W^{\prime}_{\rm SC} } 
\, . \label{WWprime}
\ee 
$W_{\rm probe} / W^{\prime}_{\rm probe} $
is the ratio of the weights in the phase space; however,
due to the particular conditions of the problem, this should effectively reflect in
the ratio of the two relative losses of gravitational weights:
\be
{W_{\rm probe} \over W^{\prime}_{\rm probe} } ~
\approx ~ {\Delta {\rm W}_g \over \Delta {\rm W}^{\prime}_g} \, .
\label{WDW}
\ee 
From~\ref{WDW}, \ref{WWprime}, \ref{WDE} and~\ref{TiTj} we 
obtain therefore:
\be
{\Delta {\rm W}^{\prime}_g \over \Delta {\rm W}_g}
~ \approx ~ 
r_T \, { \left[  1 \, + \, {\Delta_{\rm acc.} \over W_{\rm SC}} \right] 
\over
\left[  1 \, + \, r^{-1}_T {\Delta_{\rm acc.} \over W_{\rm SC}} \right] }
\label{WgT}
\, ,
\ee
where we have set $r_T \equiv T^{\prime}_c / T_c$, the ratio
of the critical temperatures of the two superconductors.
Since $\Delta_{\rm acc.} > 0$  
superconductors with higher critical temperature are expected to produce a larger
gravitational screening. The ratio
$\Delta_{\rm acc.} / W_{\rm SC}$ is not known, but it 
can be obtained by plugging in~\ref{WgT} experimental data from two
materials; in this way, this expression allows then to obtain quantitative
predictions, to be tested for any other material.

\vspace{.5cm}

\noindent
An effect analogous to the gravity screening is the impulse that seems to be
radiated by a superconductor of the same type, when it undergoes an electric discharge,
as described in~\cite{Podkletnov:2001gr}, \cite{PodkletnovModanese2002}. 
Here the singularity of the configuration, that
substitutes the accelerated motion of the superconductor,  
is produced by the high electric gradient created in a short time.
The basic idea is however the same. To be noticed is 
the instantaneous character of the modification of the geometry of space.
Sometimes it is a misleading idea to think that in quantum gravity
modifications of the geometry are propagated/mediated by gravitons,
and therefore they should obey the laws of any kind of radiation
\footnote{From a classical point of view, one may say that gravitons 
can be considered a good approximation of the description of quantum
gravity only for very weak gravitational fields. In the case of
electrodynamics things work better, because the electric field
is not charged, and the free photon remains a reasonable
approximation in a wider range of situations.}.
As pointed out is \cite{assiom-2011},
quantum mechanics is basically tachyonic. 
What is not tachyonic, and is bound
by the speed of light, is the transfer of information. In other words,
no matter of what the speed at which the quantum modification of the
geometry propagates, we can only get the information about
the results of the experiment at a speed no higher than that of light.

\newpage

\providecommand{\href}[2]{#2}\begingroup\raggedright\endgroup

\newpage

\newpage

\end{document}